\documentclass[superscriptaddress,nofootinbib,11pt,preprintnumbers]{revtex4}
\usepackage{epsfig,amsmath,amssymb}
\usepackage{color}
\usepackage{float}
\usepackage[colorlinks]{hyperref}
\usepackage{subfigure}
\usepackage[utf8]{inputenc}

\hypersetup{
    linktocpage,
     colorlinks,
     citecolor=darkgreen,
     linkcolor= darkgreen,
     urlcolor=darkgreen
}
\definecolor{darkred}{rgb}{0.5,0.0,0.0}
\definecolor{darkblue}{rgb}{0.0,0.0,0.9}
\definecolor{darkerblue}{rgb}{0.0,0.0,0.5}
\definecolor{purple}{rgb}{0.5,0.0,0.5}
\definecolor{darkgreen}{rgb}{0.0,0.5,0.0}
\definecolor{black}{rgb}{0.0,0.0,0.0}
\definecolor{brown}{rgb}{0.6,0.4,0.2}
\definecolor{newpurple}{rgb}{0.65, 0.38, 0.61}
\definecolor{newyellow}{rgb}{0.9718, 0.6093, 0.0759}
\definecolor{amber}{rgb}{1.0, 0.75, 0.0}
\definecolor{newblue}{rgb}{0.4, 0.52, 0.85}
\definecolor{newred}{rgb}{0.8524, 0.2595, 0.3294}
\definecolor{newgreen}{rgb}{0.2, 0.8, 0.2}
\definecolor{SMgreen}{rgb}{0.56, 0.69, 0.19}
\definecolor{neworange}{rgb}{0.94, 0.462, 0.162}

\definecolor{BrickRed}{rgb}{0.9,0.1,0}
\input{tcilatex}
\begin{document}

\preprint{DO-TH 15/19\quad USM-TH-339}
\title{LHC diphoton resonance at $750\,\text{GeV}$ as an indication of $%
SU(3)_L\times U(1)_X$ electroweak symmetry}
\author{A. E. C\'arcamo Hern\'andez}
\email[Electronic address:]{antonio.carcamo@usm.cl}
\affiliation{Universidad T\'{e}cnica Federico Santa Mar\'{\i}a and Centro Cient\'{\i}%
fico-Tecnol\'{o}gico de Valpara\'{\i}so\\
Casilla 110-V, Valpara\'{\i}so, Chile}
\author{Ivan Ni\v sand\v zi\' c}
\email[Electronic address:]{ivan.nisandzic@tu-dortmund.de}
\affiliation{Institut f\" ur Physik, Technische Universit\" at Dortmund, D-44221
Dortmund, Germany}
\date{\today}

\begin{abstract}
The LHC collaborations ATLAS and CMS recently reported on the excess of the
events in the diphoton final states at the invariant mass of about $750~%
\text{GeV}$. In this article we speculate on the possibility that the excess
arises from the neutral CP-even component $\phi $ of the scalar triplet $%
\Phi $ of the $SU(3)_{c}\times SU(3)_{L}\times U(1)_{X}$ $(3\text{-}3\text{-}%
1)$ model that has a $U(1)_{X}$ charge equal to $X=-1/3$ and acquires a
vacuum expectation value larger than the electroweak symmetry breaking
scale. The interactions of the scalar field $\phi$ to the photon- and
gluon-pairs are mediated by the virtual vector-like fermions which appear as
components of the anomaly-free chiral fermion representations of the $3\text{%
-}3\text{-}1$ gauge group.
\end{abstract}

\maketitle

\section{Introduction}

The experimental collaborations ATLAS and CMS recently presented the results
of the analysis of the early data obtained from the second LHC run of the
proton-proton collisions at the center-of-mass energy $\sqrt{s}=13$ TeV~\cite%
{ATLAS2015,CMS:2015dxe}. Interestingly, both experiments observed the excess
of the events with respect to the background in the diphoton final states at
the invariant mass of around $750\,\text{GeV}$. The local statistical
significance of the ATLAS (CMS) excess is about $3.9\,\sigma$ ($2.6\,\sigma$%
). The ATLAS found the signal in more than a single bin, preferring the
large width of the resonance that corresponds to about $6\%$ of its mass $%
(\simeq 45\, \text{GeV})$. This feature has not yet been confirmed by the
CMS collaboration. The available data from the second run did not reveal
additional excess of the leptons or jets at this invariant mass. While it is
possible that the reported excess is a random statistical fluctuation, if
confirmed, it would provide the first direct evidence for the physics beyond
the Standard Model (SM).

The results of many theoretical studies of the excess have been presented in
the literature in the months following the announcement. General analyses of
the excess, including surveys of several different specific model
realisations can be found in~\cite{Franceschini:2015kwy, Buttazzo:2015txu,
Gupta:2015zzs, Ellis:2015oso, Agrawal:2015dbf, Aloni:2015mxa,
Falkowski:2015swt, Csaki:2015vek}. Variety of possibilities to accommodate
the excess within the new physics models was presented in e.g.~\cite%
{Angelescu:2015uiz, DiChiara:2015vdm, Becirevic:2015fmu, Han:2015qqj,
McDermott:2015sck, Chang:2015sdy, Cao:2015pto, Dutta:2015wqh, Alves:2015jgx,
Han:2015dlp, Antipin:2015kgh, Fichet:2015vvy, Murphy:2015kag, Bauer:2015boy,
Wang:2015kuj, Petersson:2015mkr, Bellazzini:2015nxw, Demidov:2015zqn,
Bardhan:2015hcr, Ahmed:2015uqt, Pilaftsis:2015ycr, Harigaya:2015ezk,
Molinaro:2015cwg, Arun:2015ubr, Kobakhidze:2015ldh, Bian:2015kjt,
Curtin:2015jcv, No:2015bsn, Matsuzaki:2015che, Kim:2015ron, Knapen:2015dap,
Nakai:2015ptz, Backovic:2015fnp, Mambrini:2015wyu, Martinez:2015kmn,
Barducci:2015gtd, Bi:2015uqd, Heckman:2015kqk}.

Authors of the several articles~\cite%
{Franceschini:2015kwy,Buttazzo:2015txu,Angelescu:2015uiz,Ellis:2015oso,Falkowski:2015swt,Benbrik:2015fyz,Dhuria:2015ufo,Wang:2015kuj}
noted the possibility that the electrically charged and colored vector-like
fermions can be invoked for the mediation of the scalar boson interactions
to the photon and gluon pairs. In this article we identify the excess with
the scalar boson within the extended electroweak gauge group $SU(3)_L\times
U(1)_X$, that is component of the $SU(3)_L$ triplet with $U(1)_X$ charge $%
X=-1/3$. The anomaly free assignment of the fermion fields to the
representations of the $3\text{-}3\text{-}1$ group \footnote{%
In the following text we refer to the models that are based on this gauge
group as $3\text{-}3\text{-}1$ models, as the $SU(3)_c$ group factor of the
QCD remains intact.} leads to the appearance of the non-standard leptons and
quarks that are vector-like under the SM gauge group. These fermions mediate
the interactions of the scalar boson to the gluon- and photon pairs at the
loop(s) level.

\section{The Model}

The $3\text{-}3\text{-}1$ extension of the SM was first proposed in the late
seventies~\cite{Georgi:1978bv}. Several versions of the model have been
subsequently studied, see e.g.~\cite%
{Valle:1983dk,Pisano:1991ee,Montero:1992jk,Frampton:1992wt,Ng:1992st}.
Minimal versions do not include additional chiral fermion multiplets under
the $SU(3)_L\times U(1)_X$ group, beyond those that contain three
generations of the standard leptons and quarks. Many phenomenological
aspects of the model have been investigated so far. As an example, the model
can include the Peccei-Quinn symmetry, which leads to the possible solution
of the strong-CP problem~\cite{Pal:1994ba, Dias:2002gg, Dias:2003zt,
Dias:2003iq}. The studies of the models that contain sterile neutrinos in
connection with weakly interacting massive fermionic dark matter candidates
were reported in Refs.~\cite{Mizukoshi:2010ky, Dias:2010vt, Alvares:2012qv,
Cogollo:2014jia}, as well as the explorations of the fermion mass and mixing
patterns~\cite{CarcamoHernandez:2005ka, Dias:2005yh, Dias:2005jm,
Dias:2010vt, Dong:2010zu, Dong:2011vb, Dong:2011pn, Dias:2012xp,
Hernandez:2013hea, Hernandez:2013mcf, Boucenna:2014ela, Boucenna:2014dia,
Hernandez:2014vta, Hernandez:2014lpa, Vien:2014gza, Vien:2014pta,
Hernandez:2015tna, Hernandez:2015cra, Boucenna:2015zwa, Vien:2016tmh,
Hernandez:2016eod}.

We now briefly review the field content of the model and the interactions
relevant for the present discussion. The electric charge generator can be
expressed as the following linear combination 
\begin{equation}
Q=T_{3}+\beta T_{8}+XI,  \label{eq1}
\end{equation}%
where the $T_{i}$ are the generators of the $SU(3)_{L}$ group, which act on
the triplet representation via the usual Gell-Mann matrices $\lambda _{i}$,
i.e. $T_{i}=1/2\lambda _{i}$. The $X$ is the charge of the given
representation under the $U(1)_{X}$ group factor, the $I$ stands for an
identity matrix, while $\beta $ is an arbitrary real parameter.

Several versions of the $3\text{-}3\text{-}1$ models differ in the choice of
the $\beta$ parameter. The most studied versions correspond to $\beta=\pm1/%
\sqrt{3}$~\cite{Georgi:1978bv} and $\beta=\pm\sqrt{3}$~\cite{Pisano:1991ee,
Frampton:1992wt}. The standard left (right) handed quarks and leptons
are embedded into the chiral representations of the $SU(3)_L\times U(1)_X$,
i.e. as triplets (singlets) of the $SU(3)_L$ group with the corresponding
non-anomalous assignments of the $X$ charges. These representations contain
non standard fermions, which residue in the vector-like representations of
the SM gauge group. We denote the new quarks by the letter $J$ and new leptons by the
symbol $\tilde{e}$. It then follows that the cancellation of the chiral
anomalies requires that one of the quark generations residues in different
representation of the gauge group than the remaining two. As a consequence,
one obtains that the number of chiral fermion generations is a positive
integer multiple of the number of colors, which provides the theoretical
support to the observation of the existence of three generations of leptons
and quarks. For concreteness, we assign the first two generations of
left-handed quarks to the triplets of $SU(3)_L$ and the third generation to
the antitriplet representation. The assignments of the $X$-charges are
easily determined using the formula~\eqref{eq1} and requirement that the
standard leptons and quarks have correct electric charges. It turns out that
the $X$-charge of the first two generations of the left-handed triplets is
given by $X_{Q_L^{1,2}}=1/6-\beta/(2\sqrt{3})$, while for the third
generation antitriplet $X_{Q_L^{3}}=1/6+\beta/(2\sqrt{3}) $. The
corresponding $X$-charges of the right handed quarks are equal to their
electric charges, and are given as $%
X_{u_R^{1,2},d_R^{1,2},J_R^{1,2}}=2/3,-1/3,1/6-\beta \sqrt{3}/2$. The
non-standard right handed quark of the third generation carries $%
X_{J_R^{3}}=1/6+\beta \sqrt{3}/2$. All three generations of the left-handed
leptons are assigned to $SU(3)_L$ antitriplets with $X_{L_L}=-1/2-\beta/(2%
\sqrt{3})$, while the right-handed leptons are corresponding $SU(3)_L$
singlets and carry $X_{e_R,\tilde{e}_R}=-1,-1/2+\beta \sqrt{3}/2.$ Note that
the exotic fermions residue in vector-like representations of the SM gauge
group and are singlets under the $SU(2)_L$.

The scenarios with $\beta =\pm 1/\sqrt{3}$ introduce the non-standard
fermions with the non-exotic electric charges, i.e equal to the electric
charge of some standard model fermion. The options with $\beta =\pm \sqrt{3}$
involve large exotic electric charges of the new fermions, which makes these
possibilities suitable for the enhancement of the branching fraction of the
scalar resonance to the photon pairs. However, this scenario requires the
departure from the perturbative description at the scale of several $\text{%
TeV}$`s in order to remain in agreement with the measured value of the weak
mixing angle at low energies, see e.g.~\cite{Martinez:2006gb}. Other
possibilities, like $\beta =0,\pm 2/\sqrt{3}$, involve new particles with
the exotic (rational) electric charges. The electric charge conservation
forbids the decay of the lightest such particle state. The phenomenological
viability of such models would then require the detailed analysis of the
abundance of the stable exotic charged particles in the Universe's history.

We choose the value of the parameter $\beta =-1/\sqrt{3}$. The electric
charges of the vector-like quarks are $Q(J^{1,2})=2/3$ and $Q(J^{3})=-1/3$,
while the electric charges of vector-like leptons are $Q(\tilde{e}^{i})=-1$.

There are several possible choices of the scalar representations responsible
for the spontaneous symmetry breaking of the $3\text{-}3\text{-}1$ group to
the unbroken $SU(3)_{c}\times U(1)_{\text{Q}}$,~see e.g.~\cite{Diaz:2003dk}
for the detailed review. The spontaneous symmetry breaking (SSB) proceeds in
two steps. For the first step of breaking down to the SM gauge group we
choose the scalar field $\Sigma ^{ij}$ that residues in the symmetric
(sextet) representation of the $SU(3)_{L}$ and carries $X_{\Sigma }=-1/3$.
The sextet develops the non-vanishing vacuum expectation value (vev) in the
direction $\langle \Sigma ^{33}\rangle =w$, such that $w\gg v_{\text{ew}}$,
where $v_{ew}\simeq 246\,\text{GeV}$ is the vev of the standard Higgs
doublet. It turns out that this sextet does not contribute to the masses of
the fermions, since $SU(3)_{L}$ invariant Yukawa term, $\bar{\psi}_{L}\psi
_{L}^{c}\Sigma $, also requires $2X_{\psi _{L}}=X_{\Sigma }$, which is not
satisfied for any of the quark or lepton representations in the model. The
spectrum of the massive gauge bosons can be obtained from the kinetic term $%
Tr\left[ (D_{\mu }\Sigma )^{\dagger }(D^{\mu }\Sigma )\right] $ using the
expression for the covariant derivative for the sextet representation 
\begin{equation}
D_{\mu }\Sigma ^{ij}=\partial _{\mu }\Sigma ^{ij}-ig_{L}\bigg(\left( W_{\mu
}\right)^{ik}\Sigma ^{kj}+\left( W_{\mu }\right) ^{jl}\Sigma ^{li}\bigg)%
-ig_{X}X_{\mu }\delta ^{im}\Sigma ^{mj},
\end{equation}%
where $W_{\mu }=W_\mu^a T^a$ denote the gauge boson field matrix, while $X_{\mu
} $ denotes the $X$ gauge boson field. The $SU(2)_{L}\times U(1)_{Y}$
symmetry is further broken to the $U(1)_{\text{Q}}$ by two triplet
representations of the scalars, $\rho $ with $X_{\rho }=2/3$, and $\eta $
with $X_{\eta }=-1/3$. These triplets then generate the masses of the SM
fermions and $W^{\pm }$ and $Z$ gauge bosons.

We introduce the triplet $\Phi$ with the $U(1)_X$ charge $X=-1/3$ and the
vev pattern $\langle \Phi \rangle =(0,0,v_{\phi })$ to provide masses for the
exotic fermions through the Yukawa interactions 
\begin{equation}
{\small -\mathcal{L}_{\text{Y}}\supset \sum_{i=1}^{2}y_{Q}^{(i)}\overline{%
Q_{L}^{i}}\Phi J_{R}^{i}+y_{Q}^{(3)}\overline{Q_{L}^{3}}\Phi ^{\ast
}J_{R}^{3}+\sum_{i=1}^{3}y_{L}^{(i)}\overline{L_{L}^{i}}\Phi ^{\ast }\tilde{e%
}_{R}^{i}+h.c.}  \label{Lyq}
\end{equation}%
We identify the electrically neutral CP-even scalar component $\phi $ as a
candidate for the resonance at the mass equal to $750\,\text{GeV}$. The
coupling of the $\phi $ component of the triplet $\Phi$ to the vector-like fermions
is then found from the above Yukawa terms after expanding around the vacuum, $%
\phi (x)\rightarrow \phi (x)+v_{\phi }$. The scalar potential which includes
the interactions among the three $SU(3)_{L}$ scalar triplets contains a large
number of unknown couplings and is given for completeness in the Appendix~%
\ref{Scalar potential}. After the SSB there remain three physical charged
scalar bosons with masses around the $\text{TeV}$ scale and a doubly charged
scalar boson that arises from the sextet and whose mass is expected to be of
the order of the scale $w$. The contributions to the decay rate $\phi
\rightarrow \gamma \gamma$ from the loops involving charged scalars stem
from the trilinear couplings denoted by $C_{\phi S_{i}^{+(+)}S_{i}^{-(-)}}$, where $S_{i}$ labels the physical charged scalar
bosons. For example, the trilinear $C_{\phi \sigma _{1}^{++}\sigma
_{1}^{--}} $ coupling is given by $C_{\phi \sigma _{1}^{++}\sigma
_{1}^{--}}=\lambda _{14}v_{\phi }$.

\section{The resonance at $750~\text{GeV}$}

The $\phi$ boson interacts with the photon and gluon pairs via the loops of
vector-like quarks to which it couples through the Yukawa terms in the
Lagrangian~(\ref{Lyq}). The resonance is produced via gluon-gluon fusion, so
that the cross section for the proton-proton scattering into the two-photon
final state via the intermediate scalar boson $\phi $ is given in the narrow
width approximation by the formula
\begin{equation}
\sigma(pp\to\phi\to\gamma\gamma) =\frac{\pi ^{2}}{8}\frac{\Gamma (\phi
\rightarrow gg)\frac{1}{s}\int_{m_{\phi }^{2}/s}^{1}\frac{dx}{x}f_{g}(x)f_{g}%
\big(m_{\phi }^{2}/(sx)\big)\Gamma (\phi \rightarrow \gamma \gamma )}{%
m_{\phi }\Gamma _{\phi }},  \label{formula}
\end{equation}%
where $m_{\phi }\simeq 750\,\text{GeV}$ is the mass of the resonance, $%
\Gamma _{\phi } $ its total decay width and $f_{g}(x)$ denotes the parton
distribution function (pdf) of the gluon inside of the proton. We evaluate
the partial decay widths in the above formula at the leading order in QCD
and include the higher order QCD corrections by correcting the formula %
\eqref{formula} with the multiplicative factor $K^{gg}\sim 1.5$, as is
costumary.

The corresponding decay widths of the resonance are given at leading order
in QCD by 
\begin{equation}
\Gamma (\phi \rightarrow \gamma \gamma )=\frac{\alpha _{\text{em}%
}^{2}m_{\phi }^{3}}{512\pi ^{3}}\bigg\vert\sum_{i=1}^{3}\frac{N_{c}}{%
m_{J_{i}}}Q_{J_{i}}^{2}y_{Q}^{(i)}F(x_{J_{i}})+\sum_{i}\frac{1}{m_{\tilde{e}%
_{i}}}Q_{\tilde{e}_{i}}^{2}y_{L}^{(i)}F(x_{\tilde{e}_{i}})+\sum_{i}\frac{2%
\sqrt{2}C_{S_{i}}Q_{S_{i}}^{2}}{m_{\phi}^{2}}S(x_{S_{i}})\bigg\vert^{2}
\end{equation}
and 
\begin{equation}
\quad \Gamma (\phi \rightarrow gg)=\frac{\alpha _{s}^{2}m_{\phi }^{3}}{%
256\pi ^{3}}\bigg\vert\sum_{i=1,2}\frac{1}{m_{J_{i}}}y^{(i)}F(x_{i})%
\bigg\vert^{2},  \label{Gammaphoton}
\end{equation}%
where $x_{i}=4m_{i}^{2}/m_{\phi }^{2}$. The loop functions for fermion
contributions $F(x)$ and the charged scalar contribution $S(x)$ are given by
the expressions 
\begin{equation}
F(x)=2\,x\big(1+(1-x)f(x)\big),\,\quad S(x)=\left(-1+x f(x)\right) \quad 
\text{where}\quad f(x)=\big(\arcsin \sqrt{1/x}\big)^{2},
\label{loopfunctions}
\end{equation}%
valid for $x\geq1$. We use the value of the strong coupling constant $\alpha
_{s}(m_{\phi }/2)\simeq 0.1$ and the next-to-leading-order (NLO) set of pdfs
from~\cite{Martin:2009iq} (MSTW2008) at the factorization scale $\mu
=m_{\phi }$.

\begin{figure}[t]
\center
\vspace{0.7cm} \subfigure{\includegraphics[width=0.6\textwidth]{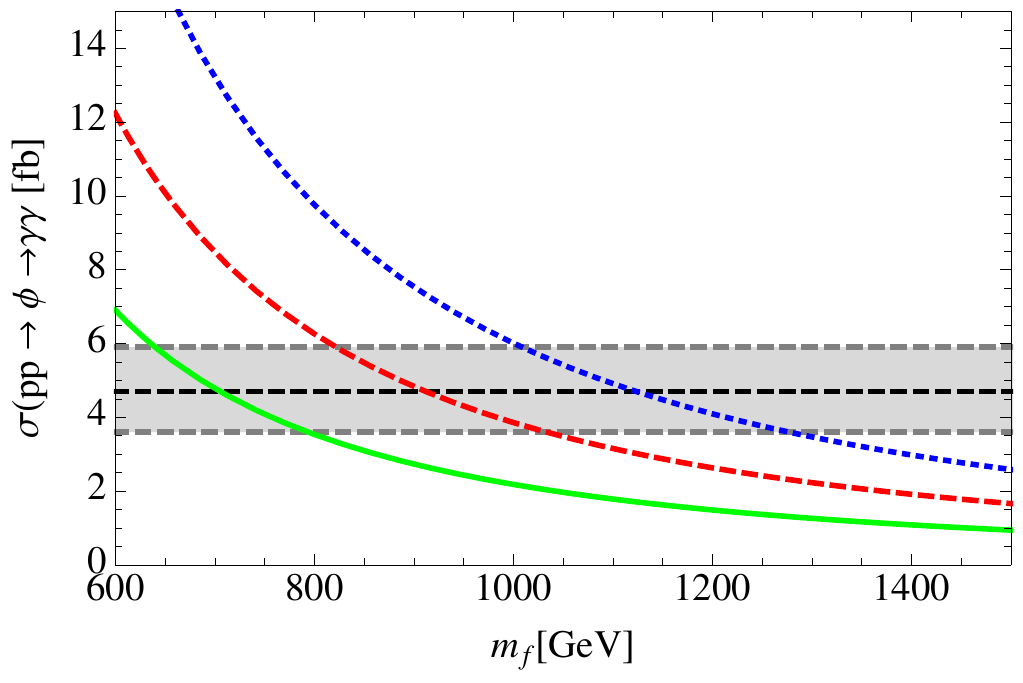}}
\caption{Total cross section of the production of the resonance $\protect%
\phi $ and in its subsequent decay into two photons at the LHC
centre-of-mass energy $13\text{ TeV}$ as a function of the common mass of
the vector-like fermions. The blue (dotted), red (dashed) and green (thick)
lines correspond to the different values of the common Yukawa couplings
equal to $2.5,2.0$ and $1.5$, respectively. The horizontal gray band
corresponds to the recent combination of the ATLAS and CMS measurements,
given in Ref. \protect\cite{Buttazzo:2016kid}.}
\label{Figure1}
\end{figure}

Given that we have $v_{\text{ew}}<v_{\phi }\ll w$ and since the couplings of
the $126\,$GeV Higgs boson are consistent the SM expectations, we
consider a benchmark scenario characterised by the absence of mixings
between the $\phi $ resonance and the remaining neutral physical scalar
fields. In addition, we assume that the $\phi $ is kinematically forbidden
to decay into charged scalar bosons. Note also that the $\phi $ boson does
not couple at the tree level to $W$ and $Z$ gauge bosons, which acquire
their masses from the $\eta $ and $\rho $ triplets. We have explicitly
checked that the contributions of the charged scalars to the diphoton rate
is subleading, so that the only relevant contribution arises from the
vectorlike fermions. For an illustration we assume that these fermions are
degenerate and show in the Fig.~\ref{Figure1} the total cross section for
the production of the $750\,\text{GeV}$ diphoton resonance at the LHC center
of mass energy $\sqrt{s}=13\text{TeV}$, as a function of the charged exotic
fermion masses $m_{F}$, and for several values of the exotic fermion Yukawa
couplings, set to be equal to $2.5,2$ and $1.5$. Keeping all the Yukawa
couplings equal and fixed to the value $1.5$, we note that the charged
exotic fermion masses cannot be higher than about $800\,\text{GeV}$, in
order to provide large enough signal cross section. For charged exotic
Yukawa couplings equal to $1.5$ and charged exotic fermion masses of $700$
GeV, we find a total cross section of $4.7$ fb and total width for the $\phi 
$ resonance of $45\,$MeV. In case that the large width of the resonance is
confirmed, the present model would be immediately excluded as the
explanation of the observed signal. This is the difficulty shared by all
(the most) weakly coupled models that aim at explaining the excess. Since
the vector-like fermions are singlets under the $SU(2)_{L}$ the decay rate $%
\Gamma (\phi \rightarrow WW)$ is also absent at one loop level. The rate $%
\Gamma (\phi \rightarrow Z\gamma )$ is suppressed with respect to the
diphoton rate by the factor $2\tan \theta _{W}=0.60$. This factor is easily
found by noticing that only the vector couplings of the $Z$ to the fermions
contributes to the corresponding amplitude, and in the limit of the heavy
scalar boson the amplitude is to a good approximation given by the amplitude
of the decay to two photons, albeit with different couplings that involve
the weak mixing angle. Furthermore, the rate $\Gamma (\phi \rightarrow ZZ)$
is even more suppressed than the rate $\Gamma (\phi \rightarrow Z\gamma )$,
since it is suppresed with respect to the diphoton rate by the factor $\tan
^{4}\theta _{W}=0.08$.

Note that the vector-like fermions may have the couplings to the standard
fermions, i.e. terms of the type $\tilde{y}^{ij}_Q\bar{Q}_L^{i}\Phi u^j_R$.
This applies also to the standard down-type quarks and charged leptons.
After the $\Phi$ develops the vev, these terms contribute to the quark
(charged lepton) mass matrices. The mixing then causes the deviations from
the unitarity of the standard Cabibbo-Kobayashi-Maskawa (CKM) matrix, and
the observable effects in the $Z$-pole and electroweak precision
observables. These effects are very tightly constrained from the available
measurements~\cite{Fajfer:2013wca, Aguilar-Saavedra:2013qpa, Allanach:2001sd}%
. In order to avoid these constraints we need to set the Yukawa couplings in
the corresponding mixing terms to some small values. This can be achieved,
at the formal level, by imposing discrete symmetry as shown in Refs.~\cite%
{Hernandez:2013hea,Hernandez:2015tna,Hernandez:2015cra,Vien:2016tmh,Hernandez:2016eod}%
. Although technically natural, setting these couplings to small values
would constitute the new flavor hierarchy problem, especially if we keep in
mind that the couplings that induce the $\phi\to\gamma\gamma$ need to be
rather large. The absence of mixings between the SM and exotic quarks will
imply that the exotic fermions will not exhibit flavor changing decays into
SM quarks and gauge (or Higgs) bosons. After being pair produced they will
decay into the standard fermions and the intermediate states of heavy gauge
bosons, which in turn decay into the pairs of the standard fermions, see
e.g. \cite{Cabarcas:2008ys}. The precise signature of the decays of the
vector-like fermions depends on details of the spectrum and other parameters
of the model. The present lower bounds from the LHC on the masses of the $%
Z^{\prime }$ gauge bosons in the $3\text{-}3\text{-}1$ models reach around $%
2.5\text{TeV}$ \cite{Salazar:2015gxa}. One can translate these bounds on the
order of magnitude of the scale $w$. The suppression of the decay rates
involving SM gauge bosons and the large masses of the nonstandard gauge
bosons then imply long-lived vector-like fermions. We plan to study the
details of the corresponding collider signatures in the future.

\section{Summary}

To summarise, we point out that the diphoton signal recently observed by the
ATLAS and CMS collaborations at the invariant mass $\sim 750\,\text{GeV}$
could arise from the $\phi$, electrically neutral CP-even component of one
of the scalar triplets representation of the $3\text{-}3\text{-}1$ model. 
Its couplings to photons and gluons are mediated by the loops that involve
exotic vector-like fermions. Such fermions appear as components of the
anomaly-free fermion representations. In order to reproduce the observed
signal, the vector like fermions need to be light (around $1\text{TeV}$) and
couple to the $\phi$ boson rather strongly. On the other hand the mixings of
the vector-like fermions to the standard chiral fermions needs to be highly
suppressed in order to remain in accordance with the precision experiments.

\textbf{Acknowledgments}. A.E.C.H was supported by DGIP internal Grant No.
111458. I. N. is supported in part by the Bundesministerium f\"ur Bildung
und Forschung (BMBF).

\begin{appendix}\section{Scalar potential}\label{Scalar potential}
The scalar potential which includes the interactions among the three $%
SU\left( 3\right) _{L}$ scalar triplets and with the scalar sextet is given
by:   
\begin{eqnarray}
V_{H} &=&\mu _{\chi }^{2}({\small \Phi }^{\dagger }{\small \Phi })+\mu
_{\eta }^{2}(\eta ^{\dagger }\eta )+\mu _{\rho }^{2}(\rho ^{\dagger }\rho
)+f_{1}\left( \eta _{i}{\small \Phi }_{j}\rho _{k}\varepsilon
^{ijk}+H.c\right) +\lambda _{1}({\small \Phi }^{\dagger }{\small \Phi })(%
{\small \Phi }^{\dagger }{\small \Phi })  \notag \\
&&+\lambda _{2}(\rho ^{\dagger }\rho )(\rho ^{\dagger }\rho )+\lambda
_{3}(\eta ^{\dagger }\eta )(\eta ^{\dagger }\eta )+\lambda _{4}({\small \Phi 
}^{\dagger }{\small \Phi })(\rho ^{\dagger }\rho )+\lambda _{5}({\small \Phi 
}^{\dagger }{\small \Phi })(\eta ^{\dagger }\eta )  \notag \\
&&+\lambda _{6}(\rho ^{\dagger }\rho )(\eta ^{\dagger }\eta )+\lambda _{7}(%
{\small \Phi }^{\dagger }\eta )(\eta ^{\dagger }{\small \Phi })+\lambda _{8}(%
{\small \Phi }^{\dagger }\rho )(\rho ^{\dagger }{\small \Phi })+\lambda
_{9}(\rho ^{\dagger }\eta )(\eta ^{\dagger }\rho )  \notag \\
&&+\mu _{\Sigma }^{2}\left( \Sigma _{ij}\Sigma ^{ij}\right) +f_{2}\left(
\eta _{i}\rho _{j}\Sigma ^{ij}+H.c\right) +f_{3}\left( {\small \Phi }%
_{i}\rho _{j}\Sigma ^{ij}+H.c\right)  \notag \\
&&+\lambda _{10}\left( \Sigma _{ij}\Sigma ^{ij}\right) \left( \Sigma
_{kl}\Sigma ^{kl}\right) +\lambda _{11}\left( \Sigma _{ij}\Sigma
^{il}\right) \left( \Sigma _{kl}\Sigma ^{jk}\right) +\lambda _{12}(\eta
^{\dagger }\eta )\left( \Sigma _{kl}\Sigma ^{kl}\right)  \notag \\
&&+\lambda _{13}(\rho ^{\dagger }\rho )\left( \Sigma _{kl}\Sigma
^{kl}\right) +\lambda _{14}({\small \Phi }^{\dagger }{\small \Phi })\left(
\Sigma _{kl}\Sigma ^{kl}\right) +\lambda _{15}\left[ ({\small \Phi }%
^{\dagger }\eta )\left( \Sigma _{kl}\Sigma ^{kl}\right) +H.c\right]  \notag
\\
&&+\lambda _{16}\eta ^{i}\Sigma _{ij}\Sigma ^{jk}\eta _{k}+\lambda _{17}\rho
^{i}\Sigma _{ij}\Sigma ^{jk}\rho _{k}+\lambda _{18}{\small \Phi }^{i}\Sigma
_{ij}\Sigma ^{jk}{\small \Phi }_{k}+\lambda _{19}\eta ^{i}\Sigma _{ij}\Sigma
^{jk}{\small \Phi }_{k}  \label{scalarpotential}
\end{eqnarray}

where the three scalar triplets and the sextet are given in terms of the components:
\begin{align}
{\small \Phi }& =%
\begin{pmatrix}
\phi _{1}^{0} \\ 
\phi _{2}^{-} \\ 
\frac{1}{\sqrt{2}}(\upsilon _{\phi }+\phi \pm i\zeta _{\phi })%
\end{pmatrix}%
,\hspace{1cm}\rho =%
\begin{pmatrix}
\rho _{1}^{+} \\ 
\frac{1}{\sqrt{2}}(\upsilon _{\rho }+\xi _{\rho }\pm i\zeta _{\rho }) \\ 
\rho _{3}^{+}%
\end{pmatrix}%
,  \notag \\
\eta & =%
\begin{pmatrix}
\frac{1}{\sqrt{2}}(\upsilon _{\eta }+\xi _{\eta }\pm i\zeta _{\eta }) \\ 
\eta _{2}^{-} \\ 
\eta _{3}^{0}%
\end{pmatrix}%
,\hspace{1cm}\Sigma =\left( 
\begin{array}{ccc}
\sigma _{1}^{0} & \sigma _{1}^{-} & \sigma _{2}^{0} \\ 
\sigma _{1}^{-} & \sigma _{1}^{--} & \sigma _{2}^{-} \\ 
\sigma _{2}^{0} & \sigma _{2}^{-} & w+\sigma _{3}^{0}%
\end{array}%
\right) .  \label{331-scalar}
\end{align}

\end{appendix}

\end{document}